\documentclass[
twocolumn,
secnumarabic,
amssymb,
amsmath,
bm,
natbib, 
nobibnotes, 
aps, 
prx,
superscriptaddress]{revtex4-2}

\usepackage{graphicx}
\PassOptionsToPackage{hyphens}{url}\usepackage{hyperref}
\usepackage{upgreek}

\usepackage{siunitx}
\DeclareSIUnit\bar{bar}

\begin{document}

\title{Demonstration of tunability of HOFI waveguides via start-to-end simulations}

\author{S. M. Mewes} 
 \email{mathis.mewesmdi@desy.de}
\affiliation{Deutsches Elektronen-Synchrotron DESY, Hamburg, Germany}
\affiliation{University of Hamburg, Hamburg, Germany}

\author{G. J. Boyle} 
\affiliation{James Cook University, Townsville, Australia}

\author{A. Ferran~Pousa} 
\affiliation{Deutsches Elektronen-Synchrotron DESY, Hamburg, Germany}

\author{R. J. Shalloo} 
\affiliation{Deutsches Elektronen-Synchrotron DESY, Hamburg, Germany}

\author{J. Osterhoff} 
\affiliation{Deutsches Elektronen-Synchrotron DESY, Hamburg, Germany}

\author{\\C. Arran} 
\affiliation{York Plasma Institute, Department of Physics, University of York, Heslington, York, United Kingdom}

\author{L. Corner} 
\affiliation{Cockcroft Institute for Accelerator Science and Technology, School of Engineering, The Quadrangle, University of Liverpool, Liverpool, United Kingdom}

\author{R. Walczak} 
\affiliation{John Adams Institute for Accelerator Science and Department of Physics, University of Oxford, Oxford, United Kingdom}

\author{S. M. Hooker} 
\affiliation{John Adams Institute for Accelerator Science and Department of Physics, University of Oxford, Oxford, United Kingdom}

\author{M. Thévenet} 
\affiliation{Deutsches Elektronen-Synchrotron DESY, Hamburg, Germany}

\date{\today}
\begin{abstract}
In recent years, hydrodynamic optical-field-ionized (HOFI) channels have emerged as a promising technique to create laser waveguides suitable for guiding tightly-focused laser pulses in a plasma, as needed for laser-plasma accelerators. While experimental advances in HOFI channels continue to be made, the underlying mechanisms and the roles of the main parameters remain largely unexplored. 
In this work, we propose a start-to-end simulation pipeline of the HOFI channel formation and the resulting guiding properties, and use it to explore the underlying physics and the tunability of HOFI channels. This approach is benchmarked against experimental measurements. HOFI channels are shown to feature excellent guiding properties over a wide range of parameters, making them a promising and tunable waveguide option for laser-plasma accelerators.
\end{abstract}


\maketitle

\section{Introduction}

Laser-plasma accelerators\,\cite{Tajima:1979,Esarey:2009} (LPAs) enable the compact acceleration of charged particles with gradients well above the $\si{\giga\electronvolt\per\meter}$ level\,\cite{Faure:2004,Geddes:2004,Mangles:2004}, orders of magnitude higher than conventional technologies.
Proof-of-principle experiments and design studies demonstrated the potential of LPA-accelerated electron beams for applications such as high-energy photon sources\,\cite{TaPhuoc:2012,Corde:2013,Wenz:2015,Zhu:2020}, free-electron lasers\,\cite{Wang:2021}, and high-energy physics\,\cite{Schroeder:2010}.
Progress in LPA performance has been strongly coupled to advances in laser technology, such as the advent of chirped pulse amplification\,\cite{Strickland:1985}, and further developments are still required to match the capabilities of conventional accelerators.
In particular, guiding of the driving laser pulse through the plasma is necessary for maintaining high accelerating gradients over multiple Rayleigh lengths, enabling energy-efficient electron acceleration in the \si{\giga\electronvolt} range\,\cite{Gonsalves:2019,Oubrerie:2022,Miao:2022}.

In recent years, Hydrodynamic Optical-Field-Ionized (HOFI) channels~\cite{Shalloo:2018} have attracted considerable attention as a promising all-optical approach to generate plasma waveguides with on-axis densities as low as $10^{17}\,\si{\per\cubic\centi\meter}$, which is required for multi-GeV electron energy gain. HOFI channels enable the guiding of tightly-focused (\SIrange[range-phrase={--}]{\sim10}{50}{\micro\metre} spot size) laser pulses, which is difficult using other guiding methods. In particular waveguides based on an electrical discharge in a gas-filled capillary~\cite{Butler:2002,Spence:2000}, require thin capillaries (easily damaged by misalignment), or complex additional plasma shaping~\cite{Gonsalves:2019}. Finally, recent experiments demonstrated the suitability of HOFI channels for plasma-based acceleration at high repetition rate~\cite{Shalloo:2019, Picksley:2020, Picksley:2020B, Miao:2020, Feder:2020, Miao:2022, Oubrerie:2022, Alejo:2022}.

The HOFI channel formation takes place in two steps. First, an ultrashort laser pulse, hereafter called the \emph{HOFI pulse}, is focused into a low-density gas to field-ionize it and thereby generate a thin plasma filament with high electron temperature.
The HOFI pulse peak intensity is chosen to be above the over-the-barrier ionization intensity ($\sim1.4\times 10^{14}\,\si{\watt\per\square\centi\meter}$) to reach full ionization at the center of the filament.
Second, due to the thermal pressure, the hot filament expands radially, producing a cylindrical blast wave with a density profile that has a dip on axis, suitable for guiding the driving laser pulse of the LPA, hereafter called the \emph{LPA pulse}.

In a further step, the power attenuation length of the channel can be increased by several orders of magnitude, to tens of metres, by ionizing the neutral gas collar surrounding the HOFI channel with a subsequent high-order Bessel pulse~\cite{Miao:2020}, or a low-order Gaussian pulse guided along the axis of the HOFI channel~\cite{Shalloo:2018b,Picksley:2020,Feder:2020}. These channels are referred to as Conditioned HOFI (CHOFI) channels.

Despite auspicious experimental realizations in the last few years, only modest progress has been made towards accurately simulating and understanding the full process of HOFI channels, from the plasma formation to the guiding of the LPA pulse.
Difficulties for such simulations are twofold. First, the system is largely \emph{multi-scale and multi-physics}: field ionization occurs over attosecond to femtosecond time scales, thermalization takes place over picoseconds, and plasma expansion and collisional ionization occur over nanoseconds. Second, the fast ionization by the HOFI pulse creates conditions for the hydrodynamic expansion that are \emph{far from equilibrium}. To the best of our knowledge, published simulation results of HOFI channel formation\,\cite{Shalloo:2018,Shalloo:2019,Picksley:2020,Oubrerie:2022} do not capture both points.

In this work, we present a multi-physics simulation pipeline covering ionization, hydrodynamic expansion and laser guiding to clarify, for the first time, the full dynamics of HOFI channels. From the properties of the HOFI pulse, the simulations enable accurate predictions---validated against experiments---of the channel formation, and exploration of the guiding properties.
In this simulation pipeline, the fast ionization from the HOFI pulse is described by particle-in-cell simulations\,\cite{Birdsall:1985} carried out with WarpX~\cite{Vay:2018}.
Then, following the success of previous authors for a wide variety of plasma problems\, \cite{bobrova2001simulations,broks2005NLTEmodel,vanDijk:2009,Bagdasarov:2017,Picksley:2020}, the subsequent expansion of the hot plasma is simulated with a hydrodynamic plasma model that does not assume Local Thermodynamic Equilibrium (LTE).
Instead, the composition of the plasma is computed using reaction rates.
This is critical to accurately capture the early dynamics of the HOFI channel formation.
Finally, the guiding properties of the resulting channel are determined using a modal solver and simplified laser propagation models in Wake-T~\cite{Benedetti:2017b,FerranPousa:2019}.

The hydrodynamic simulation model, which we have implemented as a custom model in the COMSOL Multiphysics framework, is presented in Sec.\,\ref{sec:model}. Considerations specific to HOFI channel simulations, including initial conditions and the mechanisms governing the channel formation, are presented in Sec.\,\ref{sec:HOFI_Sims}. Validation of the model against experiments is shown in Sec.\,\ref{sec:HOFI_Benchmark}. Finally, Sec.\,\ref{sec:HOFI_Guiding} discusses the implications in terms of laser guiding, and a discussion and perspectives are given in Sec.\,\ref{sec:discussion}.

\section{HYQUP Simulation Model}
\label{sec:model}

In the HYdrodynamic QUasineutral Plasma (HYQUP) model, the plasma is described as a quasi-neutral, weakly magnetized, two-temperature, reacting fluid, in an approach similar to Ref.\,\cite{broks2005NLTEmodel}. The main equations and hypotheses are presented below.

First, the plasma is described as a single fluid combining all species, whose compressible laminar mass flow is governed by the Navier-Stokes equations\,\cite{Panton2013}. These are the mass continuity equation
\begin{equation} 
    \frac{\partial \rho}{\partial t} + \Vec{\boldsymbol{\nabla}} \cdot (\rho \Vec{\boldsymbol{v}}) = 0
    \label{eqn:SIM_MassContinuity}
\end{equation}
where $\rho$ is the mass density, $t$ is the time and $\Vec{\boldsymbol{v}}$ is the flow velocity vector, and the momentum conservation equation 
\begin{equation} 
    \rho \frac{\partial \Vec{\boldsymbol{v}}}{\partial t} + \rho (\Vec{\boldsymbol{v}} \cdot \Vec{\boldsymbol{\nabla}}) \Vec{\boldsymbol{v}} = \Vec{\boldsymbol{\nabla}} \cdot ( - p \boldsymbol{I} + \boldsymbol{\tau}) + \Vec{\mathbf{F}}_{ext}
    \label{eqn:SIM_MomentumConservation}
\end{equation}
where $p$ is the pressure, $\Vec{\mathbf{F}}_{ext}$ is the sum of external forces acting on the fluid (negligible for HOFI channel formation), $\boldsymbol{I}$ is the identity matrix and $\boldsymbol{\tau}$ is the viscous stress tensor. In the model we assume a scalar viscosity $\mu$ with $\boldsymbol{\tau} = \mu[\Vec{\boldsymbol{\nabla}} \Vec{\boldsymbol{v}} + (\Vec{\boldsymbol{\nabla}} \Vec{\boldsymbol{v}})^T -\frac{2}{3}  (\Vec{\boldsymbol{\nabla}}\cdot \Vec{\boldsymbol{v}})\boldsymbol{I}]$.

Second, due to the significantly different collisional energy exchange rates, the HYQUP model uses two separate temperatures for the light particles (electrons) and the heavy particles (ions, neutrals and molecules). The difference between the temperatures can be large (e.g. laser ionization heats electrons to many $\si{\electronvolt}$, whereas the ions remain close to room temperature), and the electron-ion thermal equilibrium time is on the order of nanoseconds. The pressure in the Navier-Stokes equations is the sum of the partial pressures of these two populations. While the electrons mass is negligible to the flow, they often dominate the pressure in highly ionized plasma. The model calculates the heat transfer using separate energy conservation equations\,\cite{Bird2006} for the two particle populations. 
The heavy particle energy conservation equation reads
\begin{multline}
    \rho C_h \left( \frac{\partial T_h}{\partial t} + \Vec{\boldsymbol{v}} \cdot \Vec{\boldsymbol{\nabla}} T_h \right) 
    - \Vec{\boldsymbol{\nabla}} \cdot ( \lambda_h \Vec{\boldsymbol{\nabla}} T_h) 
    \\
    =\frac{\partial p_h}{\partial t} + \Vec{\boldsymbol{v}} \cdot \Vec{\boldsymbol{\nabla}} p_h + \boldsymbol{\tau}\!:\!\Vec{\boldsymbol{\nabla}} \Vec{\boldsymbol{v}}  + n_e\nu^\epsilon_{eh} \frac{3}{2}k_b (T_e - T_h) + Q_h
    \label{eqn:SIM_EnergyEquationHeavies}
\end{multline}
where $C_h$ is the specific heat capacity at constant pressure, $T_h$ is the temperature, $p_h$ is the partial pressure, $\lambda_h$ is the heat conductivity and $Q_h$ are additional heat sources (e.g. from ion current heating), each for heavy particle species. 
$n_{e}$ is the number density of electrons and $\nu_{eh}^\epsilon$ is the total average energy-transfer collision frequency between electrons and heavy particles. The double dot operation (:) denotes a contraction of tensors $\boldsymbol{a}:\boldsymbol{b} = \sum_{n,m} a_{n,m} b_{n,m}$. 

The electron energy conservation equation is
\begin{multline}
    \rho C_e \left( \frac{\partial T_e}{\partial t} + \Vec{\boldsymbol{v}} \cdot \Vec{\boldsymbol{\nabla}} T_e \right) 
    - \Vec{\boldsymbol{\nabla}} \cdot ( \lambda_e \Vec{\boldsymbol{\nabla}} T_e) \\
    = \frac{\partial p_e}{\partial t} + \Vec{\boldsymbol{v}} \cdot \Vec{\boldsymbol{\nabla}} p_e  - n_e\nu^\epsilon_{eh} \frac{3}{2}k_b(T_e - T_h) + Q_e
    \label{eqn:SIM_EnergyEquationElectrons}
\end{multline}
with similar notations as in Eq.\,\eqref{eqn:SIM_EnergyEquationHeavies} (subscript $e$ stands for electrons, $h$ for heavy particles). The $Q_e$ term represents additional electron heat sources e.g. due to reactions involving electrons.

For Eqs.~\eqref{eqn:SIM_EnergyEquationHeavies} and \eqref{eqn:SIM_EnergyEquationElectrons},the left-hand-side terms represent the energy change and transport by fluid flow, and heat conduction respectively. The first two right-hand-side terms represent compression heating, followed by the coupling between electrons and heavy particles, friction heating (only for heavy particles, viscosity is negligible for electrons due to their low mass), and additional heat sources, respectively.

Finally, the HYQUP model calculates the composition of particle species in the plasma mixture, which may include different gas species and chemical and ionic states. The local mass fraction of each species is tracked as they undergo reactions and diffusion, allowing for accurate representation of non-equilibrium plasma states and finite reaction rates (important for HOFI).
For all but one heavy species $\alpha$ there is a conservation equation\,\cite{Bird2006}:
\begin{equation}
    \rho \frac{\partial \omega_\alpha}{\partial t} 
    + \rho( \Vec{\boldsymbol{v}} \cdot \Vec{\boldsymbol{\nabla}}) \omega_\alpha
    + \Vec{\boldsymbol{\nabla}} \cdot \Vec{\boldsymbol{j}}_\alpha 
    = R_\alpha
    \label{eqn:SIM_SpeciesConservation}
\end{equation}
where $\omega_\alpha$ is the mass fraction, $\Vec{\boldsymbol{j}}_\alpha$ is the diffusion flux and $R_\alpha$ is the mass rate of particles being created or destroyed in reactions, all for particle species $\alpha$. The left-hand-side terms represent the change in mass fraction, transport by fluid flow and diffusion. The mass fraction of the last remaining heavy species is calculated from $\sum_\alpha \omega_\alpha = 1$, to ensure the conservation of mass. Finally the electron density distribution is calculated from the ion distributions according to quasi-neutrality.

The Equations~\eqref{eqn:SIM_MassContinuity}-\eqref{eqn:SIM_SpeciesConservation} make up the core of the HYQUP model and have been implemented as a custom model in the COMSOL Multiphysics software. Solving these differential equations requires the \emph{transport properties} $\mu$, $C_h$, $C_e$, $\lambda_h$, $\lambda_e$, $\nu_{eh}^\epsilon$ and $R_\alpha$, along with terms that describe the diffusive flux $\Vec{\boldsymbol{j}}_\alpha$. They are calculated from statistical analysis of the microscopic collisions in the plasma. Details of these calculations can be found in Appendices\,\ref{app:collisions}--\ref{app:reactions}. It is sensible to keep the model simple by reducing the choice of particle species and reactions to those significantly contributing to a given problem, e.g. ionization and recombination due to electron collisions. Further details on the treatment of the reacting mixture for a Hydrogen plasma can be found in Appendices\,\ref{app:reactions}--\ref{app:hydrogen}.

In this work, electromagnetic fields do not require consideration in the hydrodynamic part, which allows for fast simulations, taking a few minutes for one axisymmetric simulation, as described below, on a workstation (with 8-core, 3.80 GHz Intel i7-9800X CPU and 38 GB RAM). In the future, simplified electric and magnetic field solvers will be included, allowing HYQUP to have a broad range of plasma applications, similar to MHD codes such as PLASIMO\,\cite{vanDijk:2009} and FLASH\, \cite{Fryxell:2000}. Supporting non-equilibrium dynamics allows for the simulation of many plasma-based acceleration applications, by accurately describing fast processes relevant for plasma sources. In what follows, HYQUP is integrated into a simulation pipeline enabling the first experimentally benchmarked simulations of HOFI channels.

\section{HOFI channel simulation pipeline}
\label{sec:HOFI_Sims}

When simulating HOFI channel formation, three further assumptions are made with respect to the HYQUP model described in Sec.~\ref{sec:model}. First, the gas is initially uniform and the HOFI pulse focus is axisymmetric, so we assume cylindrical symmetry for the simulations. Second, we consider infinitely long plasma channels, such that the problem does not depend on the longitudinal coordinate. This assumption still permits the simulation of a finite-length HOFI channel, as long as its length is much larger than its width, see Sec.~\ref{sec:HOFI_Benchmark}. With these two assumptions, the remaining problem has a 1D radial geometry, and Neumann conditions are applied at the upper boundary (located at $r = \SI{400}{\micro\meter}$, far enough for the results to be independent of the boundary condition). Third, we only consider HOFI channels in hydrogen gas with a species mixture composed of atomic hydrogen, ionized hydrogen and electrons, unless specified otherwise. Specifics on the transport properties and reactions of the hydrogen composition, as well as an estimation of the influence of molecular hydrogen on the simulation results, can be found in Appendix\,\ref{app:hydrogen}. Hereafter, the mixture is described by the (free) electron density $n_e$ and the \emph{atomic density} $n_a$, the latter being the density of atomic nuclei, regardless of ionization, excitation or molecular states. In a purely atomic mixture this is equivalent to the heavy particle density.

\begin{figure}
	\centering
	\includegraphics[width=\columnwidth]{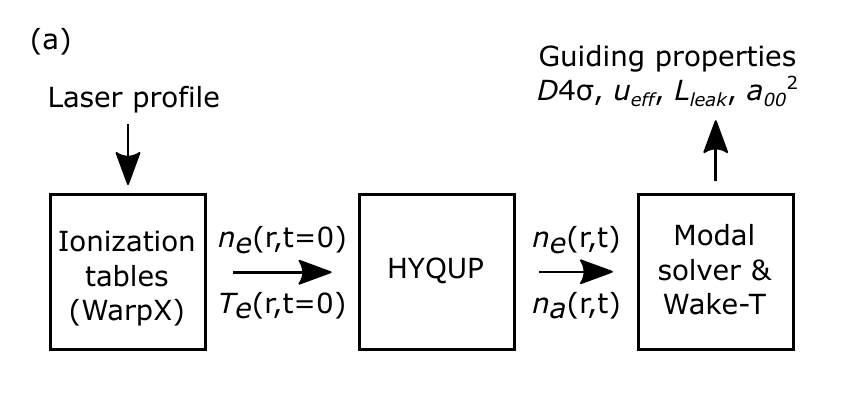}
	\includegraphics[width=\columnwidth]{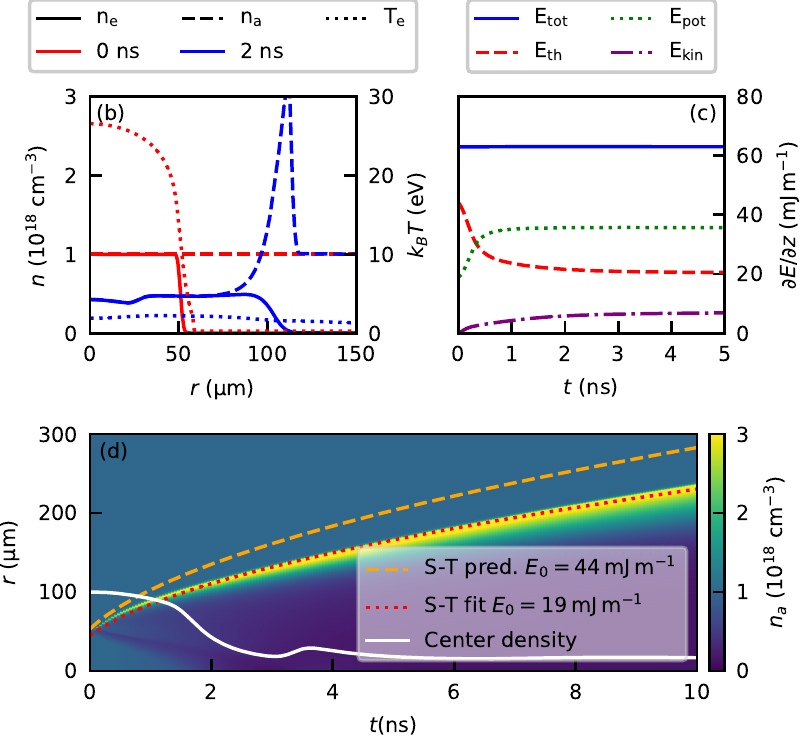}
	\caption{(a) Schematic of the multi-physics HOFI channel simulation pipeline. (b)-(c) An example simulation of a HOFI channel in an atomic hydrogen gas of initial density $n_{a,0} = 10^{18}\,\si{\per\cubic\centi\meter}$ at temperature $T_{h,0} = 300\,\si{\kelvin}$. The initial plasma channel is formed by a Gaussian HOFI pulse with central wavelength $800\,\si{\nano\meter}$, duration $30\,\si{\femto\second}$, peak intensity $10^{17}\,\si{\watt\per\square\centi\meter}$ and spot size $30\,\si{\micro\meter}$. (b) The radial distribution of electron density, atomic density and electron temperature at the beginning and after $2\,\si{\nano\second}$ of the expansion.  (c) The time evolution of the thermal $E_{th}$, potential $E_{pot}$ and kinetic energy $E_{kin}$. (d) Waterfall plot of the temporal evolution of the atomic density radial profile. The on-axis density and a comparison to the Sedov-Taylor blast theory, given by the model of Ref.\,\cite{Shalloo:2018} are also shown. Two approaches, one using Sedov-Taylor theory predictively (S-T pred.), the other fitting the Sedov-Taylor expression to the peak position obtained from HYQUP (S-T fit) are compared.}
    \label{fig:HOFI-Dynamics}
\end{figure}

The simulation pipeline we developed to obtain an accurate description of HOFI channels is shown in Fig.\,\ref{fig:HOFI-Dynamics} (a). The initial conditions for HOFI channel simulations in HYQUP consist in radially-resolved density and temperature profiles of all species (electrons, ions, neutrals). In the first step of the pipeline, starting from a cold ($\SI{300}{\kelvin}$) unionized gas, ionization properties are obtained by WarpX\,\cite{Vay:2018} simulations where field ionization is captured by the ADK model\,\cite{Ammosov:1986} with the empirical correction of Ref.\,\cite{Zhang:2014}. These simulations give, right after the passage of the HOFI pulse, the ionization fraction and the kinetic energy density of the electron population, mostly residual canonical momentum obtained at the ionization time. The electron temperature is calculated from the kinetic energy density, neglecting phenomena like ion acoustic wave excitation or collisional ionization during electron thermalization. The average free time between electron-electron collisions, estimated as the inverse of the electron collision rate (similar to Eq.\,\eqref{eq:nu_ei}), is found to be on the order of $\sim\,\si{\pico\second}$ in these conditions, justifying the simplification that thermalization happens quickly, before the hydrodynamic expansion (on the $\sim\,\si{\nano\second}$ timescale) starts. Other initial properties are trivially obtained assuming quasi-neutrality and cold ($\SI{300}{\kelvin}$) ions and neutrals.

For practical reasons, this method was used to build ionization tables: for a given set of laser polarization state, pulse duration and wavelength, the resulting ionization fraction and electron temperature were obtained from PIC simulations assuming a laser plane wave, assuming a temporal Gaussian pulse profile and scanning over a wide range of peak intensities. The resulting tabulated functions $n_e,T_e=f(I_{HOFI})$ are used to calculate the ionization properties of an arbitrary intensity profile without running a PIC simulation. This approach is appropriate when ionization-induced refraction is negligible, as is the case for this study.

In the second step, the formation of the HOFI channel by hydrodynamic expansion of the initial plasma filament over several nanoseconds is simulated using the HYQUP model, yielding detailed information about the evolution of density, mixture and temperature profiles over time. The 1D simulation space is represented by a regular grid of 14000 points and the time steps taken by the solver change adaptively from $\sim\,\si{\pico\second}$ for initialization to almost $\sim\,\si{\nano\second}$ at the end.

In the last step of the simulation pipeline, the guiding properties of the channel profiles obtained from each time iteration of the HYQUP simulation are determined through a modal solver together with guiding simulations in Wake-T\,\cite{FerranPousa:2019}. Further details of this step are described in Sec.\,\ref{sec:HOFI_Guiding}.

Figure~\ref{fig:HOFI-Dynamics} (b-c) show an example hydrodynamic simulation of the HOFI channel formation. A transversely Gaussian HOFI pulse ionizes a gas of atomic hydrogen, generating the initial temperature and density profiles shown in Fig.~\ref{fig:HOFI-Dynamics} (b). The plasma in the channel starts out fully ionized near the axis and the initial on-axis electron temperature is $\sim 26\,\si{\electronvolt}$, while the on-axis heavy-species temperature remains at the original gas temperature $300\,\si{\kelvin}$. The time taken for this temperature difference to reduce by a factor of two just by collisional energy exchange between the two particle ensembles is $\sim\SI{5}{\nano\second}$ (estimated using the energy exchange rate in Eq.\,\eqref{eqn:SIM_EnergyEquationHeavies}), which is the same time-scale as the expansion, underlining the importance of the two temperature system for accurate HOFI channel simulations. 
The initially homogeneous gas gets disturbed by a strong cylindrical blast wave, seen as a sharp spike in the atomic density moving outward over time. The resulting density profiles are suitable for guiding. The drop in on-axis electron density is solely caused by the expansion, recombination is negligible over this time scale. The bulk electron temperature drops below $2\,\si{\electronvolt}$ within the first nanosecond, which is when an LTE-based Saha equilibrium solution starts to drop below full ionization. After $5\,\si{\nano\second}$ the temperature is almost down to $1.2\,\si{\electronvolt}$ while the plasma remains almost fully ionized within the channel, conditions that could not be captured by an LTE approximation.

During the entire process the Knudsen number (an indicator to determine the type of flow) stays between 0.001 and 0.005, indicating a continuous flow regime. The Reynolds number (an indicator for turbulent flow regimes) in the bulk plasma and gas remains around $\sim1000$, making the flow non-turbulent. With both conditions fulfilled the assumption of a laminar flow is justified in these areas. In the blast wave however, the Reynolds number reaches up to almost $10^5$, potentially indicating turbulence that is not accounted for in our current model.

While the HOFI pulse deposits energy in the form of ionization and heat, the channel formation is a manifestation of kinetic energy, i.e., mass flow. Therefore, understanding the redistribution of energy during the HOFI channel formation gives considerable insight into the dynamics. This is shown on Fig.~\ref{fig:HOFI-Dynamics} (c), where the main types of energy (strictly speaking, linear energy density, per meter of channel length) are given as a function of time. These types of energy are potential energy $E_{pot} = 2 \pi \int n_{e} \epsilon_{Ion} r dr$ (proportional to the number of ionized atoms, $\epsilon_{Ion} = 13.6\,\si{\electronvolt}$ is the hydrogen ionization potential), thermal energy $E_{th} = 2 \pi \int \frac{3}{2}k_B [n_e T_e + n_h (T_h - T_{h,0})] r dr$ (excluding the initial heat of the gas), and kinetic energy $E_{kin} = 2 \pi \int \frac{1}{2} \rho v_r^2 r dr$ (accounting for the radial velocity $v_r$ of   the plasma fluid).

In half a nanosecond, collisional ionization and pressure gradients convert a fraction of the thermal energy into potential and kinetic energies, respectively. With higher initial electron temperatures, more of the thermal energy is converted into potential energy (ionizing a wider radius) in this time frame. 

The temporal evolution of the channel profile is shown in Fig.~\ref{fig:HOFI-Dynamics} (d). Note the secondary signal moving inward from the initial plasma boundary, lowering the on-axis density between $1.5 \text{ and } 2\,\si{\nano\second}$. This seems to be caused by an early pressure spike at the plasma boundary due to the collisional ionization. Although the signal is always present, it is sensitive to the exact models used, and may be reduced in realistic non-axisymmetric conditions.

A common model to describe the blast radius of the HOFI channel formation is the Sedov-Taylor (S-T) blast theory~\cite{Taylor:1950,Sedov:1959,Hutchens:2000,Shalloo:2018}, which models a blast wave starting from a kinetic energy source deposited on axis, that sweeps up all surrounding mass as it expands. To account for the finite initial size of the HOFI channel, a delay time $t_0$ is included as in Ref.\cite{Shalloo:2018}, resulting in $r(t) = \sqrt{\gamma + 1} \left( \pi m_H n_{a,0} E_{0} \right)^{-\frac{1}{4}} \sqrt{t-t_0}$ where $E_{0}$ is the initial energy deposited, $\gamma = 5/3$ is the adiabatic index and $m_H$ is the atomic mass of hydrogen. 
In Fig.\,\ref{fig:HOFI-Dynamics} (d) we demonstrate that this S-T model, which assumes the entire initial energy to be the kinetic energy of the blast, is insufficient to describe HOFI channel formation by two approaches. Firstly, we predict the S-T blast radius by assuming all the initially-deposited, linear thermal energy density is converted into the kinetic energy of the blast $E_{0} = E_{th}(t=0) = 44\,\si{\milli\joule\per\meter}$. The result shows a much wider radius than the HYQUP simulation. 
Secondly, we fit the S-T model to the density peak position obtained from HYQUP. The line fits well, but the obtained best fit energy ($E_{0} = 19.2 \pm 0.2\,\si{\milli\joule\per\meter}$) is much smaller. The difference between these two energies is not fully accounted for by the collisional ionization losses obtained from HYQUP ($\sim16\,\si{\milli\joule\per\meter}$) and the kinetic energy found in HYQUP ($\sim7\,\si{\milli\joule\per\meter}$) is also far below the fitted energy. 
Furthermore the HYQUP simulations clearly show significant amounts mass remaining inside the channel and more complicated dynamics, like collisional ionization. We conclude that this simple S-T model is not suitable for predicting the formation of HOFI channels, as neither the physical process, nor the matching energy parameter are reflected anywhere in the more detailed simulations. Furthermore, in Sec.\ref{sec:HOFI_Guiding}, we show that the guiding properties evolve mostly unrelated to the blast radius, further demonstrating the need to go beyond the S-T model to understand HOFI channels. More discussions on the initial state dependency of the hydrodynamic expansion can be found in Appendix\,\ref{app:old_fig_5}.

\begin{figure}
	\centering
	\includegraphics[width=\columnwidth]{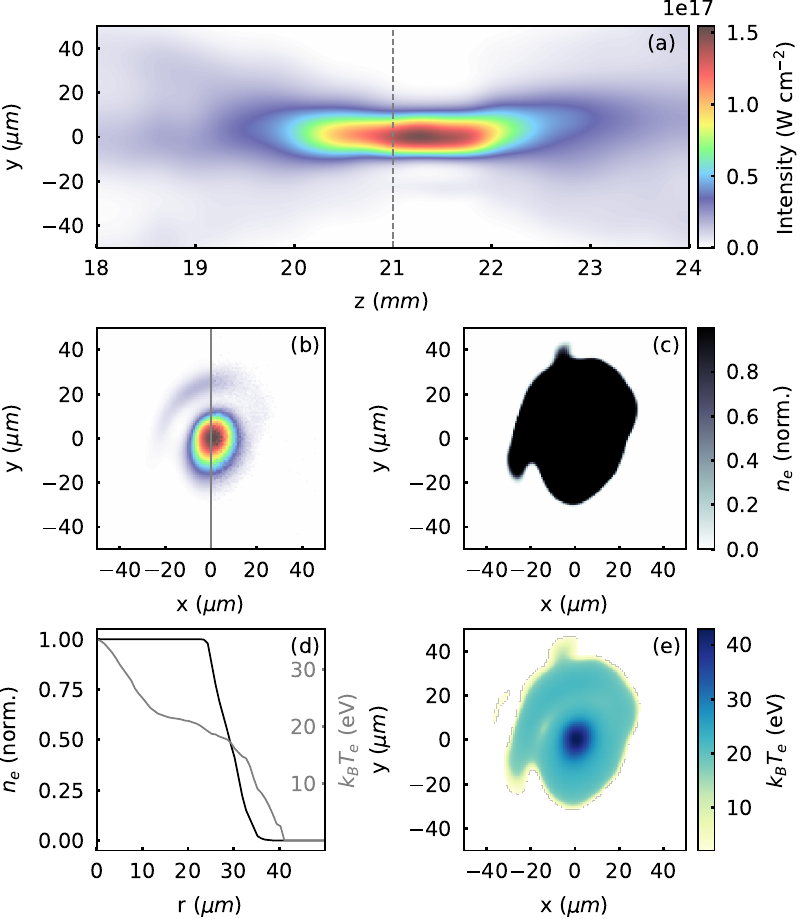}
	\caption{The determination of the initial state for the benchmark of HOFI channel simulations starts from the measured 3D HOFI pulse intensity profile shown in (a). This data set consists of a number of transverse slices as exemplified in (b) for $z = 21\,\si{\milli\meter}$. The noisy experimental data is shown in the right half and the cleaned data in the left. In (c) and (e) the ionization fraction and the electron temperature resulting from simulation of the laser ionization in this slice of the intensity profile are shown. In (d) these distributions are radially symmetrized.} \label{fig:Benchmark-Initial_Conditions}
\end{figure}

\section{Benchmark with Experiment}
\label{sec:HOFI_Benchmark}

The simulation pipeline described in Sec.\,\ref{sec:model} is benchmarked against experiments performed at the University of Oxford. Here we briefly outline the experimental setup, full details can be found in Ref.\,\cite{Shalloo:2018,Shalloo:2018b}.

The HOFI pulse was a circularly polarized Ti:sapphire laser pulse with wavelength $800\,\si{\nano\meter}$, duration $50\,\si{\femto\second}$ FWHM (intensity) and pulse energy of $25.2\,\si{\milli\joule}$, focused with a spherical lens into a short hydrogen-filled gas cell to generate the plasma filament. After a variable delay (up to a few nanoseconds), a co-propagating frequency-doubled ($\lambda=400\,\si{\nano\meter}$) pulse probed the plasma: the free electrons of the plasma and to a lesser extent the atoms of the neutral gas induce a density-dependent phase shift $\Delta\phi$, which was measured interferometrically. The measured phase shift is calibrated by using the propagation through the unionized gas as zero. The two contributions of electrons and neutral atoms could not be separated with this experimental arrangement so we compare the measured phase shift with that calculated from HYQUP results rather than attempting to deduce the electron, atomic, and ion densities from the interferometric measurements. The calculation of the phase shift is detailed in Appendix\,\ref{app:phaseshift}.

A focal scan of the HOFI pulse was taken in the experiment and used to determine the initial 3D distribution of electron density and temperature as described below. The transverse profile of the pulse was imaged every $250\,\si{\micro\meter}$ for $6\,\si{\milli\meter}$ around focus. Refraction is neglected, so the propagation of the HOFI pulse in hydrogen is approximated by vacuum propagation. As mentioned in Sec.~\ref{sec:HOFI_Sims}, the longitudinal expansion of the plasma channel is neglected, so each $z$ slice can be treated as an independent 2D transverse problem. This is in agreement with observations in Ref.\,\cite{Oubrerie:2022}, and is justified because the width of the HOFI pulse focal region is much smaller than its length (respectively $20 \text{ and } 2000\,\si{\micro\meter}$, see Fig.\,\ref{fig:Benchmark-Initial_Conditions} (a)). 

The procedure to reproduce experimental results follows several steps. First, for each slice of the HOFI pulse profile, a denoising step is applied to compensate for the low dynamic range of the measurements compared to that required for ionization calculations. The cleaned laser mode is reconstructed by decomposing the measured intensity profile into a set of Hermite-Gauss modes and filtering out the higher-order terms containing the detector noise. This procedure offers superior dynamic range to traditional noise removal methods by reconstructing the profiles from a set of analytic functions and is critical for the subsequent ionization calculations. An example slice is shown in Fig.\,\ref{fig:Benchmark-Initial_Conditions} (b), where the left side shows a cleaned profile, while the right shows noisy raw data. The collection of all slices used to reconstruct the 3D intensity profile can be seen in {Fig.\,\ref{fig:Benchmark-Initial_Conditions}~(a)}.

Second, the tables $n_e,T_e=f(I_{HOFI})$ described in Sec.\,\ref{sec:HOFI_Sims} are applied to the cleaned pulse slices to obtain the 2D electron density and temperature profiles. Such profiles are shown in Fig.\,\ref{fig:Benchmark-Initial_Conditions} (c) and (e) respectively, corresponding to the pulse intensity profile shown in Fig.\,\ref{fig:Benchmark-Initial_Conditions} (b). Importantly, the hydrogen gas was assumed fully dissociated initially, so that the HOFI pulse was ionizing atomic rather than molecular hydrogen (a discussion on molecular ionization can be found in Appendix\,\ref{app:hydrogen}). The peak intensity of the HOFI pulse significantly surpasses the ionization threshold in hydrogen. This causes the ionization profile to be almost flat-top over the high-intensity region of the HOFI pulse and sharply drop off where the intensity falls below the ionization threshold.

The electron temperature has a small bump near the laser axis, where the laser field is strong enough to excite weak plasma waves, causing plasma electrons to have a coherent longitudinal momentum $u_z$ of similar amplitude to the transverse momentum. It is unclear what fraction of $u_z$ actually decays into thermal energy. In this case $u_z$ contributes less than $\SI{5}{\percent}$ to the total thermal energy, and has a modest impact on the dynamics. When the regime is prone to strong wakefield, this could be the main source of thermal energy, governing the dynamics of the HOFI channel.

Third, the 2D electron density and temperature profiles are integrated azimuthally to give $n_e,T_e=f(r)$. This step is illustrated on Fig.\,\ref{fig:Benchmark-Initial_Conditions} (d) for the same slice. There, the smoothness of the profiles is mostly due to the azimuthal averaging: any radial cut of density in Figs.\,\ref{fig:Benchmark-Initial_Conditions} (c) and (e) would have a steeper boundary.

Fourth, the radial profiles at each slice are used as initial conditions for a hydrodynamic simulation as described in Sec.\,\ref{sec:HOFI_Sims}, giving the plasma properties resolved radially, longitudinally, and in time. 

Finally, the phase shift induced on the probe laser is calculated from the electron and neutral atom densities, as outlined in App.\,\ref{app:phaseshift}, to be compared with the experimental measurements.

\begin{figure}
	\centering
	\includegraphics[width=\columnwidth]{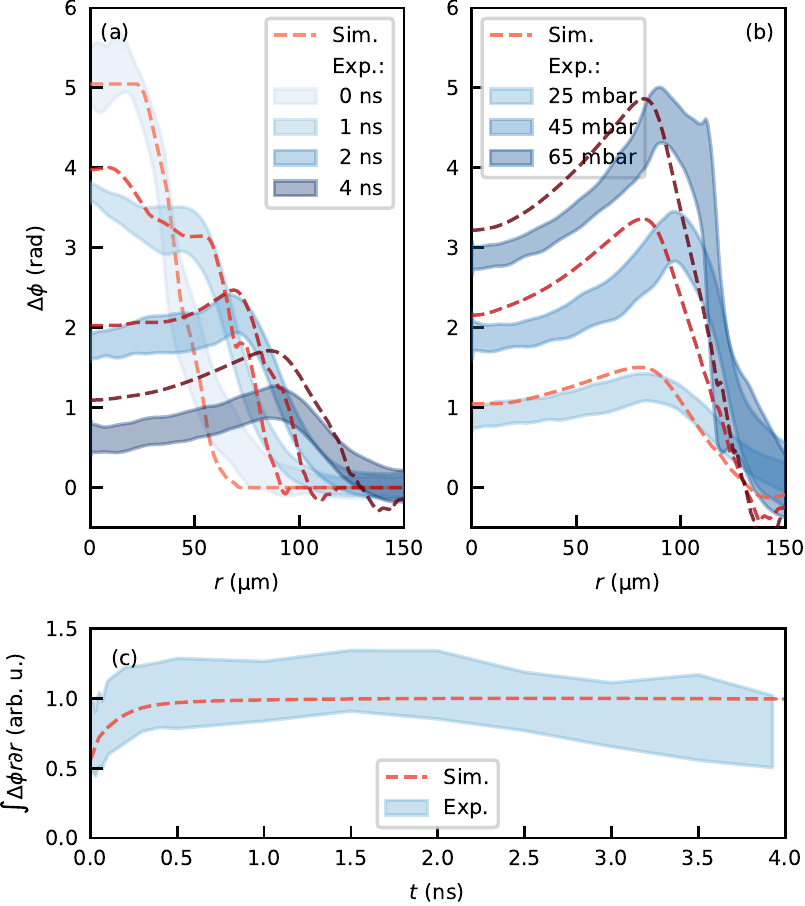}
	\caption{Comparison of experimental measurements (blue) and simulation results (red) of the phase shift induced on the probe laser, approximately proportional to the electron density. In a) a scan of the temporal evolution of the expanding HOFI channel is shown. The measurement was taken in a $2\,\si{\milli\meter}$ long gas cell, filled at a buffer pressure of $50\,\si{\milli\bar}$ of hydrogen. In (b) the profile after $3.93\,\si{\nano\second}$ of expansion is shown for different gas pressures in a $4\,\si{\milli\meter}$ long gas cell. The radial integral of the phase shift, approximately proportional to the total electron number, over the temporal scan from (a) is shown in (c). \label{fig:Benchmark} }
\end{figure}

In Fig.\,\ref{fig:Benchmark} we show comparisons between measurement and simulation results for two single-parameter scans. Due to experimental constraints, the exact positions of the gas cell relative to the laser focus is not precisely known. Thus, the gas cell position was varied on the simulated data within its experimental uncertainty of $\sim\SI{\pm1}{\milli\meter}$ to optimize the qualitative agreement between experimental and simulation results.

Figure\,\ref{fig:Benchmark} (a) shows a scan of the probe pulse delay time, capturing the temporal evolution of the channel structure, for a $2\,\si{\milli\meter}$ long gas cell operated at a $\SI{50}{\milli\bar}$ buffer pressure. In this experimental arrangement it was not possible to measure the pressure in the gas cell directly, but rather in a buffer volume nearby. To avoid systematic errors related to the pressure drop between the cell and the buffer, the initial gas density ($n_{a,0} = 2.09\times 10^{18}\si{\per\cubic\centi\meter}$) was obtained by fitting to the flat top of the first time step of the experiment, which shows the initial state immediately after ionization. The good agreement found for the initial profiles validates our procedure to obtain the initial electron density profile from the HOFI pulse profile. From the initial state of the simulation we find the total energy deposited by the HOFI pulse to be $\SI{\sim90}{\micro\joule}$, only $\sim\SI{0.36}{\percent}$ of the HOFI pulse energy. This low efficiency is not surprising for a proof-of-principle experiment and can certainly be raised by using an optimized laser system. 
Slight differences between the measured and simulated profiles at large radii could result from small changes to the initial ionization profile in the wings of the HOFI pulse intensity where the measured signal approaches the limits of the camera's dynamic range or in the conversion to and assumption of a radially symmetric expansion. The larger disagreement in signal strength observed at the longest delay may be due to experimental limitations, leading to an underestimation of the phase shift. The internal oscillations of the HOFI channel, causing the on-axis bump at $\SI{1}{\nano\second}$ delay, also seems to be observable in the experiment, although less pronounced. 
In Fig.\,\ref{fig:Benchmark} (c), we obtain the radial integral of the phase shift for each time step, which is effectively the combined phase shift of all matter in the measured volume. The simulated initial rise due to collisional ionization is in good agreement with the measurement, emphasizing the need for the finite reaction rates of the non-LTE model.

Figure\,\ref{fig:Benchmark} (b) presents a scan over the hydrogen pressure in a $4\,\si{\milli\meter}$ long gas cell, showing the state of the channel after $4\,\si{\nano\second}$. The initial condition was not measured in this scan, so the initial gas density was inferred from other measurements to be $9.05\times10^{17}\,\si{\per\cubic\centi\meter}$, $2.09\times10^{18}\,\si{\per\cubic\centi\meter}$ and $3.16\times10^{18}\,\si{\per\cubic\centi\meter}$ for the three pressures respectively. Nevertheless, a very good agreement of the magnitude of the phase shift was found, while there is a slight difference in the peak positions, that may be explained by the neglected refraction of the HOFI pulse or ionization of molecular instead of atomic hydrogen, which is further discussed in Appendix \ref{app:hydrogen}.

Overall, Fig.~\ref{fig:Benchmark} demonstrates, for the first time, excellent quantitative agreement between numerical simulations and experimental measurements of the HOFI channel formation, using only few free parameters. 
In the following section, this predictive simulation capability is used to explore the guiding properties of the resulting density profiles.
Understanding these properties and how to tune them for the optimal guiding of a laser pulse is a significant step for realizing high-performance energy-efficient LPAs.

\section{Guiding properties}
\label{sec:HOFI_Guiding}

The guiding properties of a plasma waveguide are determined by its radial electron density profile. In the case of a parabolic plasma waveguide $n(r) = n_0 + r^2/(\pi r_e w_m^4)$, where $n_0$ is the on axis electron density and $r_e$ is the classical electron radius, there exists an infinite number of bound modes (Laguerre-Gauss) \cite{Durfee:94}, the lowest order of which is characterised by a $1/e^2$ intensity radius, $w_m$, referred to as the matched spot size. 
For real plasma waveguides, in which the electron density increases radially from the axis to a peak value and then reduces finally to zero beyond the plasma region, there are no fully bound modes, but rather a set of leaky modes which propagate along the waveguide with a constant attenuation rate \cite{Durfee:94,Clark:2000}.
Typically, the lowest order mode in a real plasma waveguide used in an LPA is close to a Gaussian and will propagate with the lowest losses. Thus, here it is useful to characterize the matched spot or mode size, $w_m$ of a plasma waveguide using the D4$\upsigma$ definition for beam radius\,\cite{Siegman:1986} which matches the definition above in the case of a pure Gaussian. Additionally, it is useful to quantify the modes similarity to a purely Gaussian mode, for the identification of strongly irregular (e.g. ring-shaped) modes that are not typically used for driving an LPA.
This is quantified here by the squared amplitude $a_{00}^2$ of the fundamental Laguerre-Gauss mode contained in the guided intensity profile ($a_{00}^2\leq 1$, where $a_{00}^2=1$ corresponds to a purely Gaussian mode).
In addition to the properties of the mode, the capability of maintaining the pulse energy within the channel is an important measure for its suitability as a waveguide.
To characterize this, the relative leakage rate $R_\mathrm{leak}$ is defined as the fraction of pulse energy tunneling out of the channel per unit propagation distance.
Finally, due to the radial variation of the plasma density, an effective density value $n_{\mathrm{eff}}$ is obtained from the observed group velocity $v_g$ in the channel using the expression $n_\mathrm{eff}=(m_e\epsilon_0/e^2)w_0^2(1-v_g^2/c^2)$,
where $\omega_0$ is the laser angular frequency.

These properties are obtained from Wake-T simulations~\cite{FerranPousa:2019} of the laser pulse propagation, where its evolution in the plasma channel is modeled with an envelope solver~\cite{Benedetti:2017b}.
The laser pulse is initialized with an approximation of the matched radial intensity profile obtained by recasting the paraxial Helmholtz equation as an eigenvalue problem and numerically solving for the modes of the plasma waveguide using standard finite-difference methods.
The pulse is then propagated in the channel until convergence to the matched mode is reached.
A typical pulse duration of $\SI{30}{\femto\second}$ is assumed, and the non-linear plasma response is neglected, which is valid assuming the normalized laser amplitude parameter $a_0$ is much smaller than unity.

The guiding provided by refraction due to the plasma electrons is studied for the two extreme cases of HOFI and an optimal CHOFI for each channel.
On the one hand, if the intensity of the guided pulse is not sufficient to induce further ionization, the guiding is only provided by the basic HOFI channel, i.e. the electron density observed from the HYQUP simulation.
On the other hand, when guiding a high-power pulse (able to fully ionize the background gas) or using an additional preconditioning pulse to create a CHOFI channel, the resulting electron plasma population is instead given by the atomic density (as the plasma is fully ionized over a region much larger than the channel radius).

\begin{figure}
	\centering
	\includegraphics[width=\columnwidth]{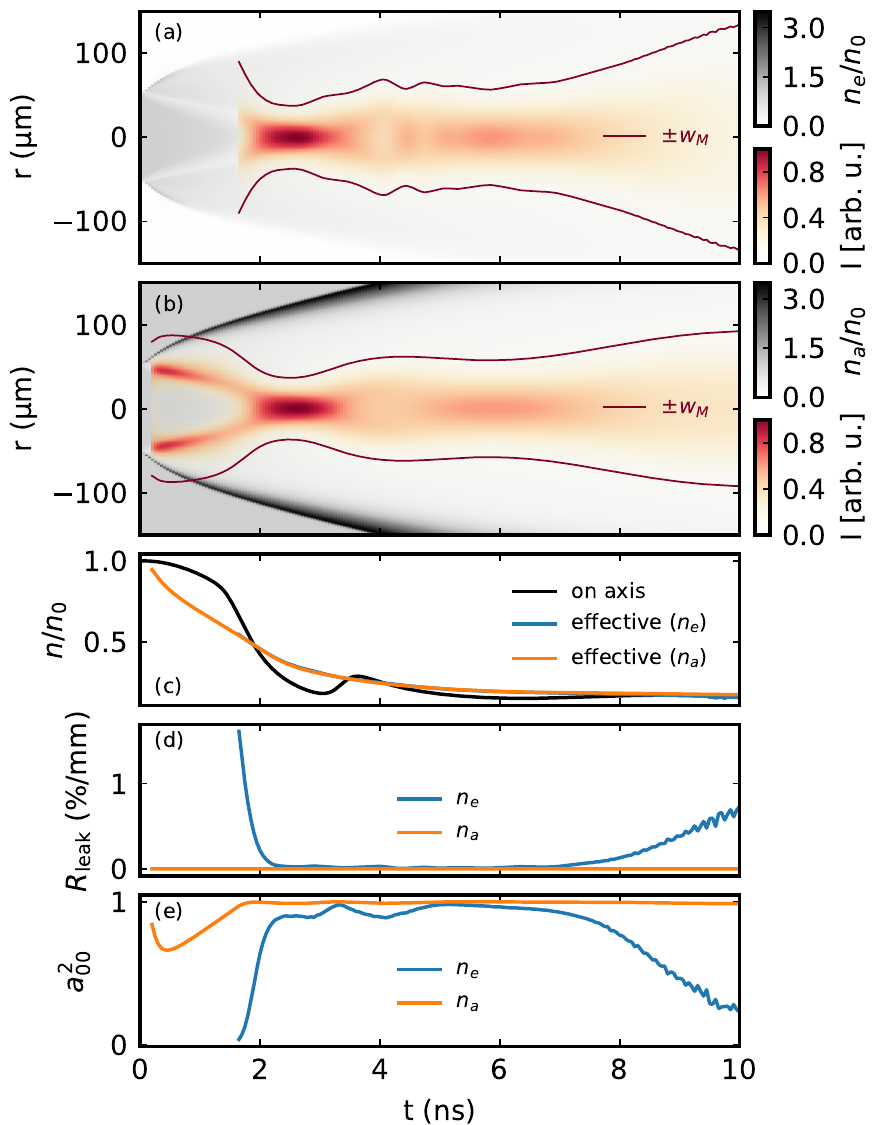}
	\caption{Evolution of the guiding properties of a HOFI channel during hydrodynamic expansion. (a) Matched mode (radial intensity profile) for the HOFI channel, (b) matched mode for a CHOFI channel that is fully ionized (up to the \SI{1}{\milli\metre} radial extent of the simulation box), (c) on-axis and effective density, (d) relative leakage rate, and (e) squared amplitude of the fundamental Laguerre-Gauss mode in the guided mode.}
    \label{fig:Laser_Scan-Guiding}
\end{figure}

The results of this simulation approach, when applied to the HOFI channel shown in Fig.~\ref{fig:HOFI-Dynamics}, are summarized in Fig.~\ref{fig:Laser_Scan-Guiding}.
A strong temporal dependence can be observed, with profiles not suitable for guiding Gaussian-like pulses when $t\lesssim \SI{2}{\nano\second}$, and with a varying matched spot size at later times.
Significant differences can also be observed between the HOFI and CHOFI channels.
While the fully-ionized atomic background in a CHOFI channel allows for negligible leakage and quasi-Gaussian modes, the initial electron density profile of the HOFI channel results in a leakage of up to $\SI{\sim1}{\%\per\milli\metre}$ and less ideal profiles, matching the guiding properties of the CHOFI channel only in narrow time intervals.
The minimal leakage in the CHOFI channel is due to the non-zero electron density outside of the channel, which prevents the pulse energy from tunnelling out.
The observed time dependence implies that the properties of the channel can be tuned by scanning the easily accessible delay between the HOFI and the LPA pulses.
In the present case, with a delay between $\SI{2}{\nano\second}$ and $\SI{7}{\nano\second}$, an effective density between $\SI{5e17}{\per\cubic\centi\meter}$ and $\SI{1e17}{\per\cubic\centi\meter}$ and a matched spot size between $\SI{40}{\micro\metre}$ and $\SI{60}{\micro\metre}$ can be used with high guiding quality.

\begin{figure}
    \centering
    \includegraphics[width=\columnwidth]{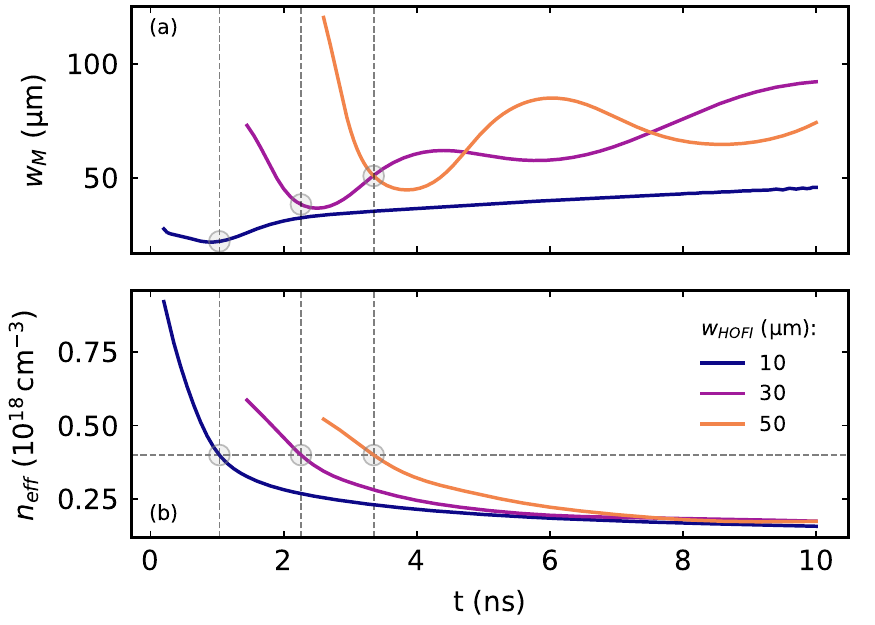}
    \caption{(a) Matched spot size and (b) effective density in HOFI channels generated by HOFI pulses with different spot size. These properties are calculated for CHOFI channels. Only the regions with high quality-guiding ($a_{00}^2>0.9$) are shown. The circles and dashed lines highlight the different time delays and corresponding matched spot sizes that can be achieved while maintaining a constant effective density of $\SI{4e17}{\per\cubic\centi\metre}$.}
    \label{fig:tunability}
\end{figure}

The high tunability of HOFI channels is further explored on Fig.~\ref{fig:tunability}.
By adjusting the spot size of the HOFI pulse and its delay with respect to the LPA pulse, the matched spot size in the channel and the effective plasma density can be independently controlled, even for a fixed gas density. As an example, for $n_{\mathrm{eff}}=\SI{4e17}{\per\cubic\centi\metre}$, $w_M$ can be scanned continuously between $20\,\si{\micro\meter}$ and $50\,\si{\micro\meter}$.
This illustrates the flexibility of HOFI and CHOFI channels, which can be adapted to maximize the performance of a wide range of laser systems for driving an LPA, without losing the high quality of the waveguide.
Some further discussion of tunability can be found in Appendix\,\ref{app:old_fig_5}.
The presented simulation method allows for an accurate and cost-effective modeling of the channel properties, therefore enabling the design and optimization of future HOFI-based LPAs.

\section{Summary}
\label{sec:discussion}

This work presents detailed simulations of HOFI channels, providing insight into the channel formation and exploring their quality over a wide range of parameters. To achieve this, HYQUP, a hydrodynamic plasma model was developed and integrated into a start-to-end simulation pipeline, which was benchmarked against experimental measurements, showing excellent agreement. Based on a single-fluid two-temperature description, HYQUP does not assume local thermodynamic equilibrium, and instead captures the evolution of the plasma composition via reactions. The guiding properties are determined by the channel shape near the axis, which is not fully captured by the reduced model of Sedov and Taylor. The simulations demonstrate that, by adjusting the HOFI pulse parameters and the delay between the HOFI pulse and the guided pulse, one can achieve large tunability of the effective density and matched radius of the waveguide independently, while maintaining a low leakage rate and preserving a Gaussian-like guided pulse. Future additions to the HYQUP model (additional gas species, interaction with boundaries, ambipolar diffusion, 2D \& 3D capabilities) will enable the simulation of HOFI channels in various conditions (e.g., suitable for ionization injection in laser-plasma accelerators), as well as its extension to simulate further hydrodynamic processes like capillary discharges.

\begin{acknowledgments}
We acknowledge support from DESY (Hamburg, Germany), a member of the Helmholtz Association HGF. We acknowledge the Funding by the Helmholtz Matter and Technologies Accelerator Research and Development Program. The authors gratefully acknowledge the Gauss Centre for Supercomputing e.V. (www.gauss-centre.eu) for funding this project by providing computing time through the John von Neumann Institute for Computing (NIC) on the GCS Supercomputer JUWELS at Jülich Supercomputing Centre (JSC).
This work was supported by the UK Science and Technology Facilities Council (STFC UK) [Grant Nos. T/J002011/1, ST/M50371X/1, ST/P002048/1]; the Engineering and Physical Sciences Research Council [Grant No. EP/ V006797/1]; and the Helmholtz Association of German Research Centres [Grant No. VHVI-503]. This material is based upon work supported by the Air Force Office of Scientific Research under Grant No. FA9550-18-1-7005.\end{acknowledgments}[floatfix]

\appendix

\section{Momentum-transfer collision frequencies}\label{app:collisions}

The average collision frequency of momentum-transfer Coulomb collisions ($\mathrm{e^- - H^+}$) and ($\mathrm{H^+ - H^+}$) are calculated numerically with 
\begin{align}
    \nu_{ei} &= \frac{4}{3}\sqrt{\frac{2\pi}{m_e}}\frac{n_e e^4 \Lambda_{ei}}{(4\pi\epsilon_0)^2\left(k_B T_e\right)^{3/2}}, \label{eq:nu_ei}\\
    \nu_{ii} &= \frac{4}{3}\sqrt{\frac{\pi}{m_i}}\frac{n_i e^4 \Lambda_{ii}}{(4\pi\epsilon_0)^2\left(k_B T_h\right)^{3/2}}, \label{eq:nu_ii}
\end{align} 
where $\epsilon_0$ is the vacuum permittivity, $k_B$ is Boltzmann's constant, $n_e$, $n_i$, $T_e$, $T_h$, $m_e$ and $m_i$ represent the electron density, ion density, electron temperature, ion temperature, electron mass and ion mass respectively, and $\Lambda_{ei}$, $\Lambda_{ii}$ are the Coulomb logarithms of electron-ion and ion-ion collisions for hydrogen. The Coulomb logarithms are calculated by
\begin{align}
    \Lambda_{ei} &= \ln{\left[\frac{3}{2\sqrt{\pi}}\sqrt{\frac{\left(4\pi\epsilon_0 k_B T_e\right)^{3}}{e^6 n_e\left(1+\frac{T_e}{T_h}\right)}}\ \right]} \label{eq:lambda_ei}, \\
    \Lambda_{ii} &= \ln{\left[\frac{3}{4\sqrt{\pi}}\sqrt{\frac{\left(4\pi\epsilon_0\right)^{3} \left( k_B T_h \right)^2 k_B T_e }{e^6 n_e\left(1+\frac{T_e}{T_h}\right)}}\ \right]}. \label{eq:lambda_ii}
\end{align} 
 A lower limit is applied to the Coulomb logarithm, i.e. $\Lambda_{ei} \rightarrow \max{\left(\Lambda_{ei},\frac{1}{2}\ln{2}\right)}$, to reduce inaccuracy for low temperatures\,\cite{Brysk_1975}.

In general, the average collision frequency for a single particle of species $\alpha$ colliding with a background of species $\beta$ (with number density $n_\beta$) is defined via
\begin{align}
    {\nu}_{\alpha\beta} = n_{\beta}\frac{4}{3}\sqrt{\frac{2\overline{\epsilon}}{m_{\alpha\beta}}}\sigma_{\alpha\beta}\left(\overline{\epsilon}\right), \label{eq:mixcolfreq}
\end{align}
where $\overline{\epsilon} = \frac{4}{\pi}k_BT_\alpha$ is the average energy, $\sigma_{\alpha\beta}$ is the corresponding cross-section, and $m_{\alpha\beta} = (m_\alpha m_\beta)/(m_\alpha+m_\beta)$ is the reduced mass. Equation~\eqref{eq:mixcolfreq} can be used to represent collisions involving neutral species given the corresponding momentum-transfer cross-sections (see Appendix~\ref{app:hydrogen}).

The energy transfer in a collision is dependent on the masses of the particles involved. The energy-transfer collision frequency, $\nu^\epsilon$, is related to the corresponding momentum-transfer collision frequency, $\nu$ via, 
\begin{align}
    \nu_{\alpha\beta}^\epsilon &\approx \frac{2m_{\alpha\beta}}{m_\alpha + m_\beta} \nu_{\alpha\beta},
\end{align}
such that the total average energy-transfer collision frequency for an electron and all heavy species is ${\nu_{eH}^\epsilon = \sum_h {\nu}_{eh}^\epsilon \approx \frac{2m_e}{m_H} \sum_h {\nu}_{eh}}$. This parameter is used in Eqs.~\eqref{eqn:SIM_EnergyEquationHeavies} and \eqref{eqn:SIM_EnergyEquationElectrons} to describe the thermal equilibration between the electrons and heavy species.

\section{Thermal conductivity and viscosity}\label{app:conductivityviscosity}

The dynamics described by Eqs.\,\eqref{eqn:SIM_MassContinuity}--\eqref{eqn:SIM_EnergyEquationElectrons} are influenced by microscopic particle interactions. The macroscopic effects on the ensemble are described by the transport properties, i.e. thermal conductivity and viscosity.
The thermal conductivity of electrons is given by\,\cite{MitchnerKruger1973},
\begin{align}
    \lambda_{e} = \frac{1}{1+\gamma_e\frac{\nu_{ei}}{\nu_{eH}}}\frac{15}{2\pi}\frac{k_B^2n_eT_e}{m_e\nu_{eH}}, \label{eq:elethermal}
\end{align}
where $\nu_{ei}$ is the average electron-ion momentum-transfer collision frequency given in Eq.~\eqref{eq:nu_ei}, and ${\nu}_{eH}=\sum_h {\nu}_{eh}$ is the total average momentum-transfer collision frequency for electrons with all heavy species, with each ${\nu}_{eh}$ defined analogously to Eq.~\eqref{eq:mixcolfreq}. The prefactor $\gamma_e$ in Eq.~\eqref{eq:elethermal} is chosen to ensure consistency with the fully ionized result \cite{MitchnerKruger1973,Braginskii1965}. 

The thermal conductivity for the mixture of heavy species is given by $\lambda_H = \sum_h{\lambda_h}$, where
\begin{align}
    \lambda_h &= \frac{1}{1+\gamma_h\frac{{\nu}_{ei}}{{\nu}_{eH}}} \frac{15}{2\pi}\frac{n_h k_B^2 T_h}{\sum_g m_{hg}{\nu}_{hg}}, \label{eq:heavythermal}
\end{align} 
where the index $g$ counts over the relevant heavy-species collisions, with each ${\nu}_{hg}$ defined analogously to Eq.~\eqref{eq:mixcolfreq}, and where the prefactor $\gamma_h$ in Eq.~\eqref{eq:heavythermal} is chosen to ensure consistency with the fully ionized result \cite{MitchnerKruger1973,Braginskii1965}.

The viscosity coefficient $\mu = \sum_h{\mu_h}$ can be deduced from the heavy species thermal conductivity simply via the Chapman-Enskog approximation\, \cite{ChapmanCowling:1970},
\begin{align}
    \mu_h = \frac{4}{15}\frac{m_h}{k_B}\lambda_h.
\end{align}

\section{Diffusion}\label{app:diffusion}
The $\Vec{\boldsymbol{j}}_\alpha$ term in Equation~\eqref{eqn:SIM_SpeciesConservation} represents the mass flux relative to the average velocity, and, assuming a mixture-averaged diffusion model, can be written as the generalized Fick's law,

\begin{equation}
    \Vec{\boldsymbol{j}}_\alpha = - \rho\omega_\alpha D_\alpha^m  \frac{\Vec{\boldsymbol{\nabla}}x_\alpha}{x_\alpha} + \rho \omega_\alpha \sum_{\beta=1}^N \omega_\beta D_\beta^m  \frac{\Vec{\boldsymbol{\nabla}}x_\beta}{x_\beta}
    \label{eqn:DiffusionFlux}
\end{equation}
where $N$ is the number of different species, $x_\alpha = \frac{\omega_\alpha}{m_\alpha}\left( \sum_{\beta=1}^N \frac{\omega_\beta}{m_\beta} \right)^{-1}$ is the mole fraction, $m_\alpha$ is the mass, and $D_\alpha^m$ is the mixture-averaged diffusion coefficient,
\begin{equation}
    D_\alpha^m = \frac{1-\omega_\alpha}{\sum_{\beta \not= \alpha} \frac{x_\beta}{D_{\alpha\beta}}},
    \label{eqn:MolecularDiffusion}
\end{equation}
all defined for species $\alpha$. The $D_{\alpha\beta}$ coefficients represent the binary Maxwell-Stefan diffusivity for species $\alpha$ and $\beta$. In this work, the binary Maxwell-Stefan diffusivities are approximated using the expression from Hartgers et al.\,\cite{Hartgers2003}, i.e., 
\begin{equation}
    D_{\alpha\beta} = \frac{p_\alpha p_\beta}{pf_{\alpha\beta}} \approx \frac{k_BT_h}{m_{\alpha\beta}\nu_{\alpha\beta}}.
\end{equation}
where $f_{\alpha\beta}= \frac{n_\alpha n_\beta}{\sum_i n_i} m_{\alpha\beta}\nu_{\alpha\beta}$ is the frictional force corresponding to the binary momentum-transfer cross-section $\sigma_{\alpha\beta}$, as in Eq.~\eqref{eq:mixcolfreq}, and once again $m_{\alpha\beta} = (m_\alpha m_\beta)/(m_\alpha+m_\beta)$ is the reduced mass. 

\section{Reactions}\label{app:reactions}

The reaction source term in the species conservation equations~\eqref{eqn:SIM_SpeciesConservation} is a sum over all considered reactions $i$, i.e.
\begin{equation}
    R_\alpha = m_\alpha \sum_i c_{i\alpha} r_i,
    \label{eqn:ReactionSourceTerm}
\end{equation} 
where $m_\alpha$ is the particle mass of species $\alpha$, $c_{i\alpha}$ is the stoichiometric number (the change of number of particles per reaction) for species $\alpha$ in reaction $i$ and $r_i$ is the rate of reaction $i$. The reaction rates 
can be written in the form,
\begin{align}
    r_i &= k_i \prod_\beta {n_\beta}^{b_{i\beta}} \label{eqn:ReactionRate} 
\end{align}
where $n_\beta$ is the number density of species $\beta$, and $b_{i\beta}$ are the stoichiometric coefficients of the reactants (number of reactants needed per reaction). The reaction constants $k_i$ are independent of the density, and are often given in the Arrhenius form (see Appendix~\ref{app:hydrogen}).

Each \textit{forward} reaction, $r^\rightarrow_i$, can have a corresponding \textit{reverse} reaction, $r^\leftarrow_i$, e.g. the collisional ionization of a neutral atom by a free electron and the recombination by a three-body collision between two electrons and an ion. The rate constant of the reverse reaction can be calculated from the forward reaction rate constant. According to the principle of \textit{detailed balance} for a mixture in chemical equilibrium, each reaction is balanced with its reverse reaction, i.e. $r^\rightarrow_i - r^\leftarrow_i = 0$. Putting Eq.\eqref{eqn:ReactionRate} in this balance we get
\begin{equation}
    k^\leftarrow_i = k^\rightarrow_i \left[ \frac{\prod_{\beta, \textrm{reactants}} {n_{\beta}}^{b_{i\beta}}}{\prod_{\beta, \textrm{products}} {n_{\beta}}^{b_{i\beta}}}\right]_{\textrm{eq}}  \equiv \frac{k^\rightarrow_i}{K_{i}^{\textrm{eq}}}  , 
    \label{eqn:ReactionBalance}
\end{equation}
where the label `eq' indicates that these are values in chemical equilibrium. The equilibrium constant $K_{i}^{\textrm{eq}}$ can then be calculated, e.g. via the Saha equations. Note that, although the reverse reaction rates are deduced from the requirements of equilibrium, in general the population densities $n_{\beta} \neq n_{\beta}^{\textrm{eq}}$, and thus $r^\rightarrow_i \neq r^\leftarrow_i$.

The reactions between the various species use or release energy to the plasma. For each reaction channel we define a heating term 
\begin{equation}
    Q_i^{\textrm{chem}} = - (r^\rightarrow_i - r^\leftarrow_i)\epsilon_{i}, 
    \label{eqn:reactionheating}
\end{equation}
where $r^\rightarrow_i$, $r^\leftarrow_i$ and $\epsilon_i$ are the forward and backward reaction rates and the reaction energy of the reaction channel $i$ (e.g. $\mathrm{H} + \mathrm{e^-} \Leftrightarrow \mathrm{H^+} + 2 \mathrm{e^-}$). In this work, all used reaction channels are based on electron collisions, where all the reaction heating contributes exclusively to the electron energy balance. The change in the average temperature on account of the changing number of particles due to reactions is included via a heat source term for each species $\alpha$, i.e.
\begin{equation}
    Q_\alpha^{\textrm{num}} = - \sum_i (r^\rightarrow_i - r^\leftarrow_i) c_{i\alpha} \frac{5}{2} k_B T_\alpha, \\
\end{equation}
where $\frac{5}{2} k_B T_\alpha$ is the enthalpy per particle of species $\alpha$. 

In the present work, unless otherwise stated, molecular hydrogen is neglected (and similarly, association/dissociation processes), such that the total heavy species density is unchanging due to reactions. The reactive heat sources in Eqs.~\eqref{eqn:SIM_EnergyEquationHeavies}--\eqref{eqn:SIM_EnergyEquationElectrons} are subsequently given by
\begin{align}
    Q_e &= \sum_{i} Q_i^{\textrm{chem}} + Q_e^{\textrm{num}}, \\
    Q_h &= 0,
\end{align}
where the sum over $i$ represents the direct and step-wise ionization processes.

\section{Hydrogen plasma components and processes}\label{app:hydrogen}

\begin{table}[ht]
    \centering
    \begin{tabular}{c|ccc}
        & \hspace{5px}$A_j\,(10^{-15}\ \si{\cubic\meter\per\second})$\hspace{5px} & \hspace{5px}$q_j$\hspace{5px} & \hspace{5px}$\epsilon_j\,(\si{\electronvolt})$\hspace{5px}  \\[2pt] \hline 
        &&&\\[-8pt]
        Direct Ionization & $7.1$ & 0.4 & 13.6\\
        Two-step Ionization & $12.4$ & 0.3 & 10.6\\
        Dissociation & $1.41$  & 2 & 4.5
    \end{tabular}
    \caption{Arrhenius parameters for the reaction rates of the hydrogen mixture. Values are taken from Broks et al.\,\cite{broks2005NLTEmodel}.}
    \label{tab:reactionparameters}
\end{table}

In this work we specifically simulate a hydrogen plasma. This is considered to be a mixture of electrons $\mathrm{e^-}$, hydrogen ions $\mathrm{H^+}$, netural hydrogen atoms $\mathrm{H}$ and molecular hydrogen $\mathrm{H_2}$, the last of which is only included when specifically stated. 

To calculate the transport properties necessary in Eqs.~\eqref{eqn:SIM_MassContinuity}--\eqref{eqn:SIM_SpeciesConservation}, as laid out in Appendices \ref{app:collisions}--\ref{app:reactions}, appropriate collision cross-sections, collision frequencies or reaction rates between the component species are needed.

The average momentum-transfer Coulomb collision frequencies ($\mathrm{e^- - H^+}$) and ($\mathrm{H^+ - H^+}$) are calculated directly via Eqs.~\eqref{eq:nu_ei}--\eqref{eq:nu_ii}. For the collisions involving neutral species we use experimental cross-section data collected from several publications. The cross-sections for collisions with neutral hydrogen atoms ($\mathrm{H^+ - H}$), ($\mathrm{H - H}$) are from Ref.\,\cite{Krstic1999}. The remaining cross-sections were retrieved from the LXCat repository\,\cite{PANCHESHNYI2012148,LXCat2017,LXCat2021}. For the collision of electrons with neutral hydrogen ($\mathrm{e^- - H}$) the Alves database\,\cite{IST-LisbonDatabase,Alves_2014} is utillized. The cross-sections for collisions with molecular hydrogen ($\mathrm{e^- - H_2}$), ($\mathrm{H^+ - H_2}$), ($\mathrm{H - H_2}$) and ($\mathrm{H_2 - H_2}$) are taken from the Phelps database\,\cite{PhelpsDatabase,Phelps1990}.

In the hydrogen mixture three reactions and their reverse reactions are considered, all based on electron collisions with heavy species. The direct ionization of hydrogen ($\mathrm{H} + \mathrm{e}^- \Leftrightarrow \mathrm{H}^+ + 2 \mathrm{e}^-$), a two-step excitation and ionization of hydrogen ($\mathrm{H} + \mathrm{e}^- \left( \Leftrightarrow \mathrm{H^*} + \mathrm{e}^- \right) \Leftrightarrow \mathrm{H}^+ + 2 \mathrm{e}^-$) and the dissociation of molecular hydrogen ($\mathrm{H}_2 + \mathrm{e}^- \Leftrightarrow 2\mathrm{H} + \mathrm{e}^-$). This selection of reactions are taken from Broks et al.\,\cite{broks2005NLTEmodel}, where they are reported as Arrhenius rates,
\begin{align}
    k_j &= A_j \left(\frac{k_B T_e}{e}\right)^{q_j} \exp\left(- \frac{\epsilon_j}{k_B T_e}\right) \label{eqn:ReactionConstant},
\end{align}
where $k_j$ is the $j$th reaction constant, and the Arrhenius parameters $A_j$, $q_j$ and $\epsilon_j$ (the reaction energy) are listed in Table\,\ref{tab:reactionparameters}.

\begin{figure}[t]
	\centering
	\includegraphics[width=\columnwidth]{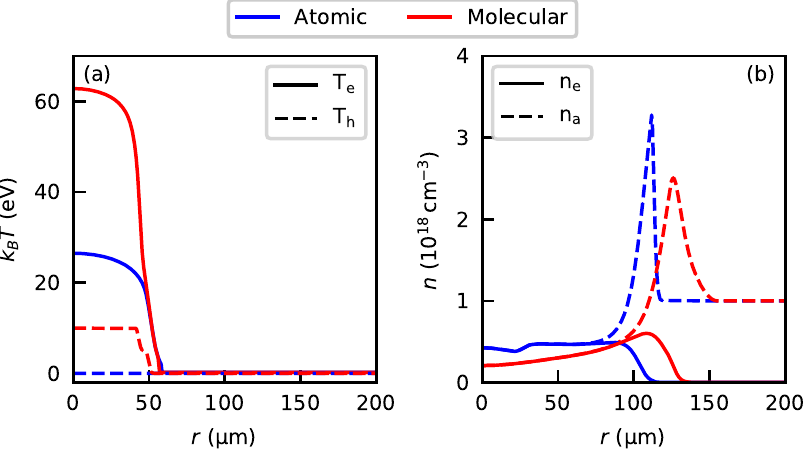}
	\caption{Effects of molecular hydrogen on the example simulation of HOFI channel formation from a Gaussian HOFI pulse shown in Sec.\,\ref{sec:HOFI_Sims}. (a) The initial temperature profiles resulting from the atomic and the rapid molecular ionization models. (b) The density profiles after $2\,\si{\nano\second}$ of expansion in the HYQUP simulation. For the molecular case the background gas is changed to initially be molecular.}
    \label{fig:H2} 
\end{figure}

For the HOFI channel simulations shown in the main sections of this paper molecular hydrogen is neglected by setting its initial mass fraction negligibly small and disabling the association and dissociation reactions. Including molecular hydrogen has two primary effects in the HOFI channel simulation pipeline. 

The first is the more complex laser ionization process of molecular hydrogen, which is still a topic of research and discussion. There is a spectrum of different ionization pathways, depending primarily on the speed of the ionization, each leading to different initial conditions for the hydrodynamic expansion. Two extreme cases are compared in Fig.\,\ref{fig:H2} (a): The slowest process fully dissociates the molecules, before ionizing the separate hydrogen atoms, leading to the initial conditions of the atomic ionization model. The fastest process is the rapid two-step ionization of the molecules, where the higher ionization energies result in a higher electron temperature. Furthermore the bare ions are left behind at the bond distance, leading to a Coulomb explosion that heats the ions. If the time between the first and second ionization step increases, the ionized molecule dissociates to a neutral hydrogen and an ion, providing a continuous spectrum of ionization energies during the process. 

The second effect of molecular hydrogen is on the hydrodynamic expansion, where the transport coefficients of the molecular background gas and the dissociation reactions slightly change the profile of the blast wave peak, but leave the blast radius and the internal channel structure essentially unchanged.

In Fig.\,\ref{fig:H2} (b) a comparison is made between the purely atomic HYQUP simulation and the molecular one, using the rapid ionization initial conditions. The additional thermal energy deposited significantly increases the blast radius and the hot ions lead to a smoother profile in the channel. This shows that further research is needed to properly account for the full range of ionization pathways and account for or even make use of them in an experimental environment.

\section{Calculation of phase shift}\label{app:phaseshift}
For the benchmark of the simulation pipeline with experimental data in Sec.\,\ref{sec:HOFI_Benchmark}, we have to calculate the phase shift induced on a probe pulse passing the plasma and neutral gas. The phase shift is obtained via

\begin{equation}
    \Delta\psi(r) = - \frac{2 \pi}{\lambda_{pr}} \int_0^L (\eta_{e}(r,z) + \eta_{n}(r,z) - 2) dz,
    \label{eq:phaseshift}
\end{equation}

where $\lambda_{pr}$ is the wavelength of the probe pulse, $L$ is the length of the gas cell and $\eta_{e}$ and $\eta_{n}$ are the refractive indices of the electrons and neutral hydrogen density respectively. The Ions contribution is omitted as negligible.
The refractive index of the neutral atomic hydrogen is calculated from a reference value of molecular hydrogen at atmospheric conditions, assuming that two neutral atoms contribute the same as one molecule. From the Lorentz-Lorenz equation\,\cite{Lorentz:1916} we derive
\begin{align}
    \eta_{n} &= \sqrt{\frac{2 K(r) + 1}{1 - K(r)}},\label{eq:refractiveIndexH}\\
    K(r) &= \frac{n_{n}(r)}{2 n_{atm}} \frac{{\eta_{atm}}^2 - 1}{{\eta_{atm}}^2 + 2} ,\label{eq:refractiveIndexHK}
\end{align}
where $n_n$ is the neutral hydrogen number density, $n_{atm} = 2.69\times10^{19}\si{\per\cubic\centi\meter}$ is the atmospheric number density at $\SI{0}{\celsius}$ and $\eta_{atm}$ is the refractive index of molecular hydrogen at atmospheric number density\,\cite{Peck:1977}
\begin{equation}
    \eta_{atm} = 1 + \frac{0.0148956}{180.7 - \left(\frac{1 [\si{\micro\meter}]}{\lambda_{pr}}\right)^2} + \frac{0.0049037}{92 - \left(\frac{1 [\si{\micro\meter}]}{\lambda_{pr}}\right)^2}.\label{eq:refractiveIndexAirReference}
\end{equation}

The refractive index of free plasma electrons can be calculated from the critical plasma density $n_{cr}$ using
\begin{align}
    \eta_e &= \sqrt{1 - \frac{n_e}{n_{cr}}},\label{eq:refractiveIndexE}\\
    n_{cr} &= m_e \epsilon_0 \left(\frac{2 \pi c_0}{e\lambda_{pr}}\right)^2,\label{eq:criticalDensity}
\end{align}
where $m_e$ is the electron mass, $c_0$ is the vacuum speed of light and $e$ is the elemental charge.
Finally the phase shift is calibrated by the propagation in undisturbed gas to match with the measurement. For this we use
\begin{equation}
    \Delta\phi(r) = \Delta\psi(r) - \Delta\psi(r_{max}),\label{eq:refractiveIndexCalibrated}
\end{equation}
with the phase shift at the boundary of the simulation box $r_{max}$, where the gas remains undisturbed through the simulation.

\section{Initial state dependency}\label{app:old_fig_5}
\begin{table}[htb]
    \centering
    \begin{tabular}{c|ccccc}
        Case & $w_{HOFI}$ ($\si{\micro\meter}$) & $\lambda_{HOFI}$ ($\si{\nano\meter}$) & $E_{tot}$  ($\si{\milli\joule\per\meter}$) & $E_{th}/E_{tot}$ (\%) \\[2pt] \hline
        &&& \\[-8pt]
        1& $36.9$ &400& 43.8&39 \\
        2& $25.2$ &800&43.8&71 \\
        3& $18.6$ &1200&43.8&84 \\
        4& $52.1$ &400&87.5&39\\
        5& $35.6$ &800&87.5&71 \\
        6& $26.3$ &1200&87.5&84 \\
    \end{tabular}
    \caption{Properties of the HOFI pulse (beam waist $w_{HOFI}$ and wavelength $\lambda_{HOFI}$) and corresponding total energy $E_{tot}$ and fraction of thermal energy $E_{th}/E_{tot}$ for the simulations in Fig.\,\ref{fig:Energy_Scan}a) (cases 1-3) and Fig.\,\ref{fig:Energy_Scan}b) (cases 4-6). The fraction of potential energy is simply given by $E_{pot}/E_{tot}=1-E_{th}/E_{tot}$.}
    \label{tab:energies}
\end{table}
\begin{figure}[ht]
	\centering
	\includegraphics[width=\columnwidth]{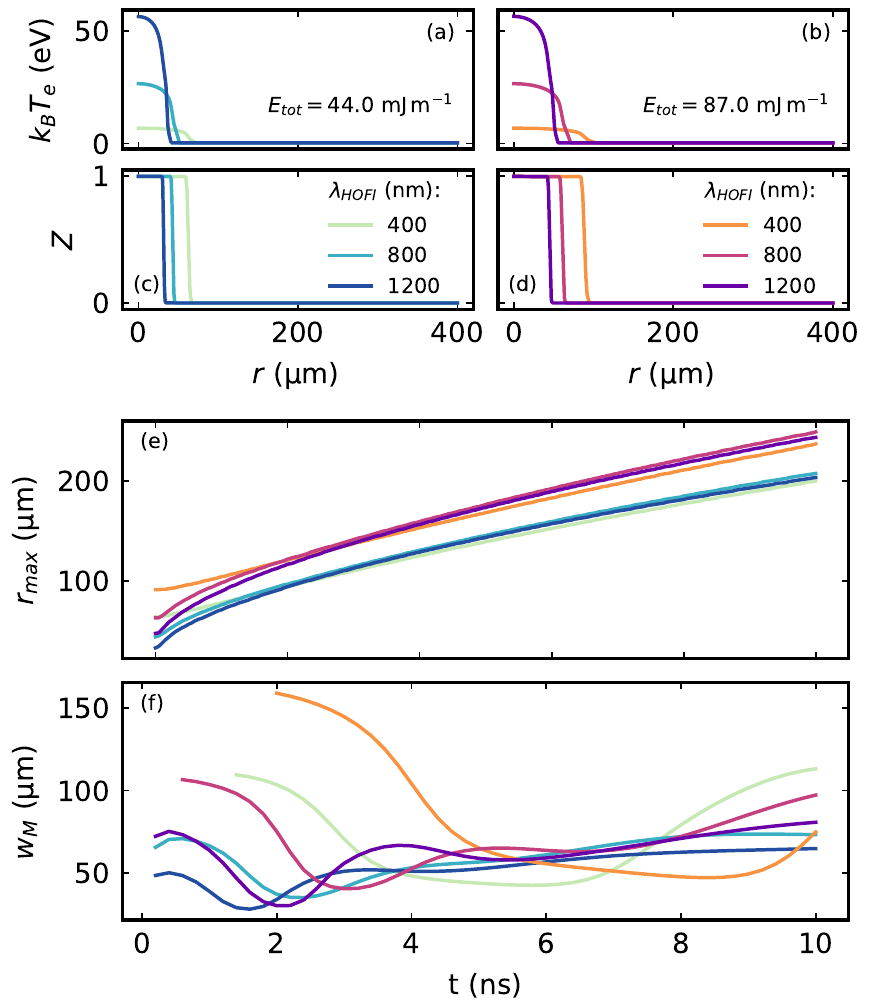}
	\caption{Influence of the initial energy distribution on the HOFI channel formation. (a) and (c) Radial profile of the initial electron temperature and ionization fraction, respectively, for three configurations of the HOFI pulse. The parameters (beam waist and central wavelength) were chosen to vary the balance between thermal and potential energy while keeping the sum (total energy) constant to $43.7\,\si{\milli\joule\per\meter}$. (b) and (d) Same as (a) and (c) for a total energy of $87.4\,\si{\milli\joule\per\meter}$. Details are shown in Table\,\ref{tab:energies}. (e) Evolution of the atomic density peak position for the 6 simulations in (a)-(d). (f) Evolution of the matched spot size for the same simulations.}
    \label{fig:Energy_Scan}
\end{figure}

In Sec.\,\ref{sec:HOFI_Sims} we found the Sedov-Taylor model to be insufficient to properly describe the expansion of a HOFI channel. This raises the question if there is a different way to predict the blast radius from the initial energy deposited, which can be found based on the potential energy $E_{pot}$, the thermal energy $E_{th}$ or their combined total $E_{tot}=E_{pot}+E_{th}$. The thermal energy strongly depends on the wavelength of the HOFI pulse $\lambda_{HOFI}$ ($E_{th}\propto \lambda_{HOFI}^2$)~\,\cite{Schroeder:2014}, so the balance between thermal and potential energies can be adjusted by changing the HOFI pulse width and wavelength, while keeping the total deposited energy constant. This is illustrated on Fig.\,\ref{fig:Energy_Scan} (a)-(d), showing initial conditions of plasma channels for two values of the total energy with three variations of the energy distribution each, detailed in Table\,\ref{tab:energies}. The simulated peak position dynamics of these six cases, shown on Fig.\,\ref{fig:Energy_Scan}~(e), are approximately the same for a given total energy, after $\sim2\,\si{\nano\second}$. This is suggesting a simple model, similarly to S-T, should still be viable, using the total energy deposited by the HOFI pulse. 

Nevertheless, the main interest lies not with the peak position, but the guiding properties of the HOFI channels formed around the center. The evolution of the matched spot size in the CHOFI channels formed in the six cases is shown in Fig.\,\ref{fig:Energy_Scan}~(f), where the lines are limited to sections where the quality of the waveguide is good (negligible losses, and Gaussian mode ($a_{00}^2>0.9$). The guiding properties of the channel are again largely independent of the peak position shown above, demonstrating that a simplified blast model cannot be used to infer the guiding properties as they vary strongly with the different distribution of initial energy deposition. On the one hand, it demonstrates the wide range of tunability that simple changes of the HOFI pulse can provide, without impeding on the quality of the waveguide. On the other hand, this shows that detailed knowledge of the HOFI pulse is critical to predict the guiding properties. It also opens up the question of how instability of the HOFI pulse could disturb the operational stability of HOFI channels over many shots, which warrants further studies.

\end{document}